# CAAD: Computer Architecture for Autonomous Driving

Shaoshan Liu, Jie Tang, Zhe Zhang, and Jean-Luc Gaudiot, *Fellow, IEEE*

**ABSTRACT**
We describe the computing tasks involved in autonomous driving, examine existing autonomous driving computing platform implementations. To enable autonomous driving, the computing stack needs to simultaneously provide high performance, low power consumption, and low thermal dissipation, at low cost. We discuss possible approaches to design computing platforms that will meet these needs.

## 1. INTRODUCTION

An autonomous vehicle must be capable of sensing its environment and safely navigating without human input. Indeed, the US Department of Transportation's National Highway Traffic Safety Administration (NHTSA) has formally defined five different levels of autonomous driving [1]:

- Level 0: the driver completely controls the vehicle at all times; the vehicle is not autonomous at all.

- Level 1: semi-autonomous; most functions are controlled by the driver, but some functions such as braking can be done automatically by the vehicle.

- Level 2: the driver is disengaged from physically operating the vehicle by having no contact with the steering wheel and foot pedals. This means that at least two functions, cruise control and lane-centering, are automated.

- Level 3: there is still a driver who may completely shift safety-critical functions to the vehicle and is not required to monitor the situation as closely as for the lower levels.

- Level 4: the vehicle performs all safety-critical functions for the entire trip, and the driver is not expected to control the vehicle at any time since this vehicle would control all functions from start to stop, including all parking functions.

Levels 3 and 4 autonomous vehicles must sense their surroundings by using multiple sensors, including LiDAR, GPS, IMU, cameras, *etc*. Based on the sensor inputs, they need to be able to localize themselves, and in real-time, make decisions about how to navigate within the perceived environment. Due to the enormous amount of sensor data and the high complexity of the computation pipeline, autonomous driving places extremely high demands in terms of computing power and electrical power consumption. Existing designs often require equipping an autonomous car with multiple servers, each with multiple high-end CPUs and GPUs. These designs come with several problems: first, the costs are extremely high, thus making autonomy unaffordable to the general public. Second, power supply and heat dissipation become a problem as this setup consumes thousands of Watts, consequently placing high demands on the vehicle's power system.

We explore computer architecture techniques for autonomous driving. First, we introduce the tasks involved in current LiDAR-based autonomous driving. Second, we explore how vision-based autonomous driving, a rising paradigm for autonomous driving, is different from the LiDAR-based counterpart. Then, we look at existing system implementations for autonomous driving. Next, considering different computing resources, including CPU, GPU, FPGA, and DSP, we attempt to identify the most suitable computing resource for each task. Based on the results of running autonomous driving tasks on a heterogeneous ARM Mobile SoC, we propose a system architecture for autonomous driving, which is modular, secure, dynamic, energy-efficient, and is capable of delivering high levels of computing performance.

## 2. TASKS IN AUTONOMOUS DRIVING

Autonomous Driving is a highly complex system that consists of many different tasks. As shown in Figure 1, in order to achieve autonomous operation in urban situations with unpredictable traffic, several real-time systems must interoperate, including sensor processing, perception, localization, planning and control [2]. Note that existing successful implementations of autonomous driving are mostly LiDAR-based: they rely heavily on LiDAR for mapping, localization, and obstacle avoidance, while other sensors are used for peripheral functions [3, 4].

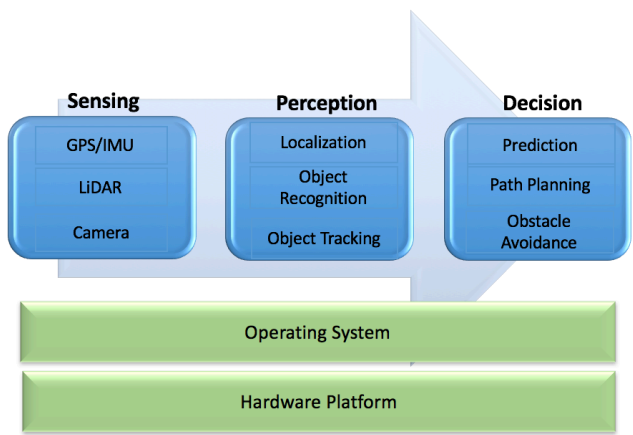

Figure 1: Tasks in Autonomous Driving: consisting of three main stages, sensing, perception, and decision.

## 2.1 Sensing

Normally, an autonomous vehicle consists of several major sensors. Indeed, since each type of sensor presents advantages and drawbacks, in autonomous vehicles, the data from multiple sensors must be combined for increased reliability and safety. They can include the following:

### 2.1.1 GPS and Inertial Measurement Unit (IMU)

The GPS/IMU system helps the autonomous vehicle localize itself by reporting both inertial updates and a global position estimate at a high rate. GPS is a fairly accurate localization sensor, but its update rate is slow, at about only 10 Hz, and thus not capable of providing real-time updates. Conversely, an IMU's accuracy degrades with time, and thus cannot be relied upon to provide reliable position updates over long periods of time. However, an IMU can provide updates more frequently, at or higher than 200 Hz to satisfy the real-time requirement. Assuming a vehicle traveling at 60 miles per hour, the traveled distance is less than 0.2 meters between two position updates, (this means that the worst case localization error is less than 0.2 meters).

By combining both GPS and IMU, we can provide accurate and real-time updates for vehicle localization. Nonetheless, we cannot rely on this sole combination for localization for three reasons: 1.) its accuracy is only about one meter; 2.) the GPS signal has multipath problems, meaning that the signal may bounce off buildings, introducing more noise; 3.) GPS requires an unobstructed view of the sky and would thus not work in environments such as tunnels.

### 2.1.2 LiDAR

LiDAR is used for mapping, localization, and obstacle avoidance. It works by bouncing a laser beam off of surfaces and measures the reflection time to determine distance. Due to its high accuracy, it is used as the main sensor in most autonomous vehicle implementations. LiDAR can be used to produce high-definition maps, to localize a moving vehicle against high-definition maps, to detect obstacles ahead, *etc*. Normally, a LiDAR unit, such as Velodyne 64-beam laser, rotates at 10 Hz and takes about 1.3 million readings per second. There are two main problems with LiDAR: 1.) when there are many suspended particles in the air, such as rain drops and dust, the measurements may be extremely noisy. 2.) a 64-beam LiDAR unit is quite costly.

### 2.1.3 Camera

Cameras are mostly used for object recognition and object tracking tasks such as lane detection, traffic light detection, and pedestrian detection, *etc*. To enhance autonomous vehicle safety, existing implementations usually mount eight or more 1080p cameras around the car, such that we can use cameras to detect, recognize, and track objects in front of, behind, and on both sides of the vehicle. These cameras usually run at 60 Hz, and, when combined, would generate around 1.8 GB of raw data per second.

### 2.1.4 Radar and Sonar

The radar and sonar system is mostly used as the last line of defense in obstacle avoidance. The data generated by radar and sonar shows the distance to the nearest object in front of the vehicle's path. Once we detect that an object is close ahead, there may be a danger of a collision, then the autonomous vehicle should apply the brakes or turn to avoid the obstacle. Therefore, the data generated by radar and sonar does not require much processing and usually is fed directly to the control processor, and thus not through the main computation pipeline, to implement such "urgent" functions as swerving, applying the brakes, or pre-tensioning the seatbelts.

## 2.2 Perception

After getting sensor data, we feed the data into the perception stage to understand the vehicle's environment. The three main tasks in autonomous driving perception are localization, object detection, and object tracking.

### 2.2.1 Localization

Localization is a sensor-fusion process, such that GPS/IMU, and LiDAR data can be used to generate a high-resolution infrared reflectance ground map. To localize a moving vehicle relative to these maps, we could apply a particle filter method to correlate the LiDAR measurements with the map [10]. The particle filter method has been demonstrated to achieve real-time localization with 10-centimeter accuracy and to be effective in urban environments. However, the high cost of LiDAR could limit its wide application.

### 2.2.2 Object Detection

In recent years, however, we have seen the rapid development of vision-based Deep Learning technology, which achieves significant object detection and tracking accuracy [7]. Convolution Neural Network (CNN) is a type of Deep Neural Network that is widely used in object recognition tasks. A general CNN evaluation pipeline usually consists of the following layers: 1.) The Convolution Layer which contains different filters to extract different features from the input image. Each filter contains a set of "learnable" parameters that will be derived after the training stage. 2.) The Activation Layer which decides whether to activate the target neuron or not. 3.) The Pooling Layer which reduces the spatial size of the representation to reduce the number of parameters and consequently the computation in the network. 4.) The Fully Connected Layer where neurons have full connections to all activations in the previous layer. The convolution layer is often the most computation-intensive layer in a CNN.

### 2.2.3 Object Tracking

Object tracking refers to the automatic estimation of the trajectory of an object as it moves. After the object to track is identified using object recognition techniques, the goal of object tracking is to automatically track the trajectory of the object subsequently. This technology can be used to track nearby moving vehicles as well as people crossing the road to ensure that the current vehicle does not collide with these moving objects. In recent years, deep learning techniques have demonstrated advantages in object tracking compared to conventional computer vision techniques [11]. Specifically, by using auxiliary natural images, a stacked Auto-Encoder can be trained offline to learn generic image features that are more robust against variations in viewpoints and vehicle positions. Then, the offline trained model can be applied for online tracking.

## 2.3 Decision

Based on the understanding of the vehicle's environment, the decision stage can generate a safe and efficient action plan in real-time. The tasks in the decision stage mostly involve probabilistic processes and Markov chains.

### 2.3.1 Prediction

One of the main challenges for human drivers when navigating through traffic is to cope with the possible actions of other drivers which directly influence their own driving strategy. This is especially true when there are multiple lanes on the road or when the vehicle is at a traffic change point [12]. To make sure that the vehicle travels safely in these environments, the decision unit generates predictions of nearby vehicles, and decides on an action plan based on these predictions. To predict actions of other vehicles, one can generate a stochastic model of the reachable position sets of the other traffic participants, and associate these reachable sets with probability distributions.

### 2.3.2 Path Planning

Planning the path of an autonomous, agile vehicle in a dynamic environment is a very complex problem, especially when the vehicle is required to use its full maneuvering capabilities. A brute force approach would be to search all possible paths and utilize a cost function to identify the best path. However, the brute force approach would require enormous computation resources and may be unable to deliver navigation plans in real-time. In order to circumvent the computational complexity of deterministic, complete algorithms, probabilistic planners have been utilized to provide effective real-time path planning [13].

### 2.3.3 Obstacle Avoidance

As safety is the paramount concern in autonomous driving, at least two levels of obstacle avoidance mechanisms need to be deployed to ensure that the vehicle will not collide with obstacles. The first level is proactive, and is based on traffic predictions [14]. At runtime, the traffic prediction mechanism generates measures like time to collision or predicted minimum distance, and based on this information, the obstacle avoidance mechanism is triggered to perform local path re-planning. If the proactive mechanism fails, the second-level, the reactive mechanism, using radar data, will take over. Once the radar detects an obstacle, it will override the current control to avoid the obstacles.

## 3. VISION-BASED AUTONOMOUS DRIVING

LiDAR is capable of producing over a million data points per second with a range up to 200 meters. However, it is very costly (a high-end LiDAR sensor costs over tens of thousands of dollars). We thus explore an affordable yet promising alternative, vision-based autonomous driving.

## 3.1 LiDAR *vs.* Vision Localization

The localization method in LiDAR-based systems heavily utilizes a particle filter [3], while vision-based localization utilizes visual odometry techniques [6]. These two different approaches are required to handle the vastly different types of sensor data. The point clouds generated by LiDAR provide a "shape description" of the environment, but it is hard to differentiate individual points. By using a particle filter, the system compares a specific observed shape against the known map to reduce uncertainty. In contrast, for vision-based localization, the observations are processed through a full pipeline of image processing to extract salient points and the salient points' descriptions, which is known as feature detection and descriptor generation. This allows us to uniquely identify each point and apply these salient points to directly compute the current position.

## 3.2 Vision-Based Localization Pipeline

In detail, vision-based localization undergoes the following simplified pipeline: 1.) by triangulating stereo image pairs, we first obtain a disparity map which can be used to derive depth information for each point. 2.) by matching salient features between successive stereo image frames, we can establish correlations between feature points in different frames. We can then estimate the motion between the past two frames. 3.) Also, by comparing the salient features against those in the known map, we can also derive the current position of the vehicle.

## 3.3 Impact on Computing

Compared to a LiDAR-based approach, a vision-based approach introduces several highly parallel data processing stages, including feature extraction, disparity map generation, optical flow, feature match, Gaussian Blur, *etc*. These sensor data processing stages heavily utilize vector computations and each task usually has a short processing pipeline, which means that these workloads are best suited for DSPs. In contrast, a LiDAR-based approach heavily utilizes the Iterative Closest Point (ICP) algorithm [15], which is an iterative process that is hard to parallelize, and thus more efficiently executed on a sequential CPU.

## 4. EXISTING IMPLEMENTATIONS

To understand the main points in autonomous driving computing platforms, we look at an existing computation hardware implementation of a level 4 autonomous car from a leading autonomous driving company. Then, to understand how the chip makers attempt to solve these problems, we look at the existing autonomous driving computation solutions provided by different chip makers.

## 4.1 Computing Platform Implementation

Our interaction with a leading autonomous driving company (name withheld by request) has led us to understand that their current computing platform consists of two compute boxes, each equipped with an Intel Xeon E5 processor and four to eight Nvidia K80 GPU accelerators, connected with a PCI-E bus. At its peak performance, the CPU (which consists of 12 cores), is capable of delivering 400 GOPS/s, consumes 400 W of power. Each GPU is capable of 8TOPS/s, while consuming 300 W of power. Combining everything together, the whole system is able to deliver 64.5 TOPS/s at about 3000 W. The compute box is connected to twelve high-definition cameras around the vehicle, for object detection and object tracking tasks. A LiDAR unit is mounted on top of the vehicle for vehicle localization as well as some obstacle avoidance functions. A second compute box performs exactly the same tasks and is used for reliability: in case the first box fails, the second box can immediately take over. In the worst case, when both boxes run at their peak, this would mean over 5000 W of power consumption which would consequently generate enormous amount of heat. Also, each box costs 20 ~ 30 thousand dollars, making the whole solution unaffordable to average consumers.

## 4.2 Existing Processing Solutions

We examine some existing computing solutions targeted for autonomous driving.

### 4.2.1 GPU-Based Solutions

The Nvidia PX platform is the current leading GPU-based solution for autonomous driving. Each PX 2 consists of two Tegra SoCs and two Pascal graphics processors. Each GPU has its own dedicated memory, as well as specialized instructions for Deep Neural Network acceleration. To deliver high throughput, each Tegra connects directly to the Pascal GPU using a PCI-E Gen 2 x4 bus (total bandwidth: 4.0 GB/s). In addition, the dual CPU-GPU cluster is connected over Gigabit Ethernet, delivering 70 Gigabits per second. With optimized I/O architecture and DNN acceleration, each PX2 is able to perform 24 trillion deep-learning calculations every second. This means that, when running AlexNet deep learning workloads, it is capable of processing 2,800 images/s.

### 4.2.2 DSP-Based Solutions

Texas Instruments' TDA provides a DSP-based solution for autonomous driving. A TDA2x SoC consists of two floating-point C66x DSP cores and four fully programmable Vision Accelerators, which are designed for vision processing functions. The Vision Accelerators provide eight-fold acceleration on vision tasks compared to an ARM Cortex-15 CPU, while consuming less power. Similarly, CEVA XM4 is another DSP-based autonomous driving computing solution. It is designed for computer vision tasks on video streams. The main benefit for using CEVA-XM4 is energy-efficiency, which requires less than 30mW for a 1080p video at 30 frames per second.

### 4.2.3 FPGA-Based Solutions

Altera's Cyclone V SoC is one FPGA-based autonomous driving solution which has been used in Audi products. Altera's FPGAs are optimized for sensor fusion, combining data from multiple sensors in the vehicle for highly reliable object detection. Similarly, Zynq UltraScale MPSoC is also designed for autonomous driving tasks. When running Convolution Neural Network tasks, it achieves 14 images/sec/Watt, which outperforms the Tesla K40 GPU (4 images/sec/Watt). Also, for object tracking tasks, it reaches 60 fps in a live 1080p video stream.

### 4.2.4 ASIC-Based Solutions

MobilEye EyeQ5 is a leading ASIC-based solution for autonomous driving. EyeQ5 features heterogeneous, fully programmable accelerators, where each of the four accelerator types in the chip are optimized for their own family of algorithms, including computer-vision, signal-processing, and machine-learning tasks. This diversity of accelerator architectures enables applications to save both computational time and energy by using the most suitable core for every task. To enable system expansion with multiple EyeQ5 devices, EyeQ5 implements two PCI-E ports for inter-processor communication.

# 5. COMPUTER ARCHITECTURE DESIGN EXPLORATION

We attempt to develop some initial understandings of the following questions: 1.) what computing units are best suited for what kind of workloads 2.) if we considered an extreme, would a mobile processor be sufficient to perform the tasks in autonomous driving, and 3.) how to design an efficient computing platform for autonomous driving?

## 5.1 Matching Workloads to Computing Units

We seek to understand which computing units are best fitted to convolution and feature extraction workloads, which are the most computation-intensive workloads in autonomous driving scenarios. We conducted experiments on an off-the-shelf ARM mobile SoC consisting of a four-core CPU, a GPU, as well as a DSP, the detailed specifications can be found in [8]. To study the performance and energy consumption of this heterogeneous platform, we implemented and optimized feature extraction and convolution tasks on CPU, GPU, and DSP, and measured chip-level energy consumption.

First, we implemented a convolution layer, which is commonly used, and is the most computation-intensive stage, in object recognition and object tracking tasks. The left side of Figure 2 summarizes the performance and energy consumption results: when running on the CPU, each convolution takes about 8 ms to complete, consuming 20 mJ; when running on the DSP, each convolution takes 5 ms to complete, consuming 7.5 mJ; when running on a GPU, each convolution takes only 2 ms to complete, consuming only 4.5 mJ. These results confirm that GPU is the most efficient computing unit for convolution tasks, both in performance and in energy consumption.

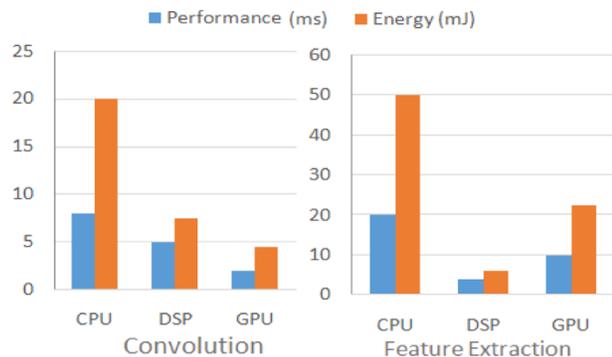

Figure 2: Convolution and Feature Extraction Performance and Energy: DSP is best suited for feature extraction, and GPU is best suited for convolution.

Next, we implemented feature extraction, which generates feature points for the localization stage, and this is the most computation expensive task in the localization pipeline. The right side of Figure 2 summarizes the performance and energy consumption results: when running on a CPU, each feature extraction task takes about 20 ms to complete, consuming 50 mJ; when running on a GPU, each convolution takes 10 ms to complete, consuming 22.5 mJ; when running on a DSP, each convolution takes only 4 ms to complete, consuming only 6 mJ. These results confirm that DSP is the most efficient computing unit for feature processing tasks, both in performance and in energy consumption. Note that we did not implement other tasks in autonomous driving, such as localization, planning, obstacle avoidance *etc*. on GPUs and DSPs as these tasks are control-heavy and would not efficiently execute on GPUs and DSPs.

## 5.2 Autonomous Driving on Mobile Processor

We seek to explore the edges of the envelope and understand how well an autonomous driving system could perform on the aforementioned ARM mobile SoC. Figure 3 shows the vision-based autonomous driving system we implemented on this mobile SoC. We utilize the DSP for sensor data processing tasks, such as feature extraction and optical flow; we use GPU for deep learning tasks, such as object recognition; we use two CPU threads for localization tasks to localize the vehicle at real-time; we use one CPU thread for real-time path planning and we use one CPU thread for obstacle avoidance. Note that multiple CPU threads can run on the same CPU core if a CPU core is not fully utilized.

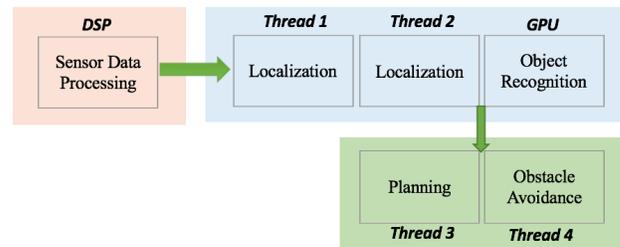

Figure 3: Autonomous Navigation System on Mobile SoC: we fully utilize the heterogeneous computing platform to achieve performance and energy efficiency.

Surprisingly, it turns out that the performance was quite impressive when we ran this system on the ARM Mobile SoC. The localization pipeline is able to process 25 images per second, almost keeping up with image generation at 30 images per second. The deep learning pipeline is capable of performing 2 to 3 object recognition tasks per second. The planning and control pipeline is designed to plan a path within 6 ms. When running this full system, the SoC consumes 11 W on average. With this system, we were able to drive the vehicle at around 5 miles per hour without any loss of localization, quite a remarkable feat, considering that this ran on a mobile SoC. With more computing resources, the system should be capable of processing more data and allowing the vehicle to move at a higher speed, eventually satisfying the needs of a production-level autonomous driving system.

## 5.3 Design of Computing Platform

The reason why we could deliver this performance on an ARM mobile SoC is that we fully utilized the heterogeneous computing resources of the system and used the best suited computing unit for each task so as to achieve best possible performance and energy efficiency. However, there is a downside as well: we could not fit all the tasks into such a system, for example, object tracking, change lane prediction, cross-road traffic prediction, *etc.* In addition, we need for the autonomous driving system to have the capability to upload raw sensor data and processed data to the cloud but the amount of data is so large that it would take all of the available network bandwidth.

The aforementioned functions, object tracking, change lane prediction, cross-road traffic prediction, data uploading *etc.* are not needed all the time. For example, the object tracking task is triggered by the object recognition task and the traffic prediction task is triggered by the object tracking task. The data uploading task is not needed all the time either since uploading data in batches usually improves throughput and reduces bandwidth usage. If we designed an ASIC chip for each of these tasks, it would be a waste of chip area, but an FPGA would be a perfect fit for these tasks. We could have one FPGA chip in the system and have these tasks time-share the FPGA. It has been demonstrated that using Partial-Reconfiguration techniques [5], an FPGA soft core could be changed within less than a few milliseconds, making time-sharing possible in real-time.

Figure 4: Computing Stack for Autonomous Driving: consisting of application, operating system, runtime, and computing layers.

In Figure 4, we show our proposed computing stack for autonomous driving. At the level of the computing platform layer, we have an SoC architecture consisting of an I/O subsystem that interacts with the front-end sensors; a DSP to pre-process the image stream to extract features; a GPU to perform object recognition and some other deep learning tasks; a multi-core CPU for planning, control, and interaction tasks; an FPGA that can be dynamically reconfigured and time-shared for data compression and uploading, object tracking, and traffic prediction, *etc*. These computing and I/O components communicate through shared memory. On top of the computing platform layer, we have a run-time layer to map different workloads to the heterogeneous computing units through OpenCL, and to schedule different tasks at runtime with a run-time execution engine. On top of the Run-Time Layer, we have an Operating Systems Layer utilizing Robot Operating System (ROS) design principles [9], which is a distributed system consisting of multiple ROS nodes, each encapsulating a task in autonomous driving.

## 5.4 Discussion

At PerceptIn, we have implemented and shipped products with the aforementioned autonomous driving computing stack, which provides several benefits: 1.) it is modular: more ROS nodes can be added if more functions are required 2.) it is secure: ROS nodes provide a good isolation mechanism to prevent nodes from impacting each other 3.) it is highly dynamic: the run-time layer can schedule tasks for max throughput, lowest latency, or lowest energy consumption 4.) it can deliver high performance: each heterogeneous computing unit is used for the most suitable task to achieve highest performance 5.) it is energy-efficient: we can use the most energy-efficient computing unit for each task, for example, a DSP for feature extraction.

## 6. CONCLUSIONS

Existing computing solutions for Level 4 autonomous driving often consume thousands of Watts, dissipate enormous amounts of heat, and cost tens of thousands of dollars. These power, heat, and cost barriers thus make autonomous driving technologies difficult to transfer to the general public. We proposed and developed an autonomous driving computing architecture and software stack that is modular, secure, dynamic, high-performance, and energy-efficient. Our prototype system on an ARM Mobile SoC consumes 11 W on average and is able to drive a mobile vehicle at 5 miles per hour. With more computing resources, the system will be able to process more data and will eventually satisfy the need of a production-level autonomous driving system.

## 7. ACKNOWLEDGMENTS

This work is partly supported by the National Science Foundation under Grant No. XPS-1439165. Any opinions, findings, and conclusions or recommendations expressed in this material are those of the authors and do not necessarily reflect the views of NSF.

Dr. Shaoshan Liu is the co-founder of PerceptIn. He attended UC Irvine for his undergraduate and graduate studies and obtained a Ph.D. in Computer Engineering in 2010. His research focuses on Computer Architecture, Big Data Platforms, Deep Learning Infrastructure, and Robotics. He has over eight years of industry experience: before co-founding PerceptIn, he was with Baidu USA, where he led the Autonomous Driving Systems team. Before joining Baidu USA, he worked on Big Data platforms at LinkedIn, Operating Systems kernel at Microsoft, Reconfigurable Computing at Microsoft Research, GPU Computing at INRIA (France), Runtime Systems at Intel Research, and Hardware at Broadcom. Email: shaoshan.liu@perceptin.io

Dr. Jie Tang is the corresponding author and she is currently an associate professor in the School of Computer Science and Engineering of South China University of Technology, Guangzhou, China. Before joining SCUT, Dr. Tang was a post-doctoral researcher at the University of California, Riverside and Clarkson University from Dec. 2013 to Aug. 2015. She received the B.E. from the University of Defense Technology in 2006, and the Ph.D. degree from the Beijing Institute of Technology in 2012, both in Computer Science. From 2009 to 2011, she was a visiting researcher at the PArallel Systems and Computer Architecture Lab at the University of California, Irvine, USA. Email: cstangjie@scut.edu.cn

Dr. Zhe Zhang is the co-founder of PerceptIn. He received the Bachelor's degree in Automation from Tsinghua University, Beijing, China in 2005. He received a PhD degree in Robotics from State University of New York (SUNY) at Stony Brook, NY, USA in 2009. His PhD was on vision based robotic 3D mapping and localization. From 2009 to 2014, Zhe was with Microsoft Robotics working on mapping, localization, navigation, and self-recharge solutions on a prototype consumer robot. In May 2014, Zhe joined Magic Leap Mountain View office and was leading sparse/dense mapping and pose tracking efforts. Email: zhe.zhang@perceptin.io

Dr. Jean-Luc Gaudiot received the Diplôme d'Ingénieur from ESIEE, Paris, France in 1976 and the M.S. and Ph.D. degrees in Computer Science from UCLA in 1977 and 1982, respectively. He is currently Professor in the Electrical Engineering and Computer Science Department at UC, Irvine. Prior to joining UCI in 2002, he was Professor of Electrical Engineering at the University of Southern California since 1982. His research interests include multithreaded architectures, fault-tolerant multiprocessors, and implementation of reconfigurable architectures. He has published over 250 journal and conference papers. His research has been sponsored by NSF, DoE, and DARPA, as well as a number of industrial companies. He has served the community in various positions and was just elected to the presidency of the IEEE Computer Society for 2017. E-mail: gaudiot@uci.edu